\documentclass[a4paper,11pt]{article}
\usepackage{jcappub} 

\arxivnumber{2510.11030} 
\title{\boldmath Resonant W and Z Boson Production in FSRQ Jets: Implications for Diffuse Neutrino Fluxes}

\author[a,b,1]{J.-H. Ha\note{Corresponding author.}}
\author[c]{and I. Alikhanov}
\affiliation[a]{Korea Space Weather Center, Korea AeroSpace Administration,\\
198-6 Gwideok-ro, Jeju 63025, Republic of Korea}
\affiliation[b]{Korea Astronomy and Space Science Institute,\\
776 Daedeok-daero, Yuseong-gu, Daejeon 34055, Republic of Korea}
\affiliation[c]{North-Caucasus Center for Mathematical Research, North-Caucasus Federal University,\\
Stavropol 355017, Russian Federation}

\emailAdd{hjhspace223@gmail.com}

\abstract{Blazars, particularly Flat Spectrum Radio Quasars (FSRQs), are well-known for their ability to accelerate a substantial population of electrons and positrons, as inferred from multiwavelength radiation observations. 
Therefore, these astrophysical objects are promising candidates for studying high-energy electron--positron interactions, such as the production of $W^{\pm}$ and $Z$ bosons. 
In this work, we explore the implications of electron--positron annihilation processes in the jet environments of FSRQs, focusing on the resonant production of electroweak bosons and their potential contribution to the diffuse neutrino flux. 
By modeling the electron distribution in the jet of the FSRQ 3C~279 during a flaring state, we calculate the reaction rates for $W^{\pm}$ and $Z$ bosons and estimate the resulting diffuse fluxes from the cosmological population of FSRQs.
We incorporate the FSRQ luminosity function and its redshift evolution to account for the population distribution across cosmic time, finding that the differential flux contribution exhibits a pronounced peak at redshift $z \sim 1$. 
While the expected fluxes remain well below the detection thresholds of current neutrino observatories such as IceCube, KM3NeT, or Baikal-GVD, the flux from $Z$ boson production within the jet blob is many orders of magnitude smaller than the total diffuse astrophysical neutrino flux. These results provide a theoretical benchmark for the role of Standard Model electroweak processes in extreme astrophysical environments, highlighting the interplay between particle physics and astrophysics, and illustrating that even extremely rare high-energy interactions can leave a subtle, theoretically meaningful imprint on the diffuse astrophysical neutrino background.}

\begin{document}
\maketitle
\flushbottom

\section{Introduction}
\label{sec:s1}
Blazars are a subclass of active galactic nuclei characterized by bipolar relativistic jets aligned closely with our line of sight. 
This jet orientation, combined with Doppler boosting, causes jet emission to dominate observations across nearly all wavelengths. 
Blazars are divided into two categories: flat spectrum radio quasars (FSRQs), which exhibit strong broad emission lines, and BL~Lac objects, which have weak or absent lines \cite[e.g.,][]{abrahamyan2023}. 
The broad spectral energy distributions (SEDs) observed in blazars are produced by multiple radiation mechanisms. 
The lower-frequency component mainly arises from synchrotron emission by nonthermal electrons (including positrons) accelerated within the jet. 
The higher-energy component can be generated by Compton scattering between these nonthermal electrons and photons. 
In particular, synchrotron radiation provides seed photons for the Compton scattering process, which is known as Synchrotron Self-Compton (SSC). 
Alternatively, Compton scattering can involve lower-energy seed photons entering the jet externally (External Compton, EC) from sources such as the accretion disk \cite[e.g.,][]{dermer1992, dermer1993}, broad-line region \cite[e.g.,][]{sikora1994,blandford1995,ghisellini1996}, or dust torus \cite[e.g.,][]{kataoka1999, blazejowski2000}.

Blazars are characterized by stochastic variability, including quiescent and flaring states \cite[e.g.,][]{finke2014,finke2015}. 
For example, the FSRQ 3C 279 alternates between these states over periods ranging from a few days to several weeks \cite[e.g.,][]{hayashida2012,hayashida2015}. 
Multiwavelength modeling reveals significant differences in the electron distribution function depending on the states \cite[e.g.,][]{lewis2019}. 
During flares, the electron distribution exhibits a higher maximum Lorentz factor and a flatter high-energy spectral slope, indicating more efficient particle acceleration likely due to enhanced turbulence or shock strength. 
In contrast, the quiescent state shows a steeper high-energy cutoff, reflecting dominant radiative cooling and weaker acceleration. 
These changes in the electron distribution shape directly influence the observed SEDs, shifting both the synchrotron and inverse-Compton peaks in frequency and flux.

The mechanism responsible for electron acceleration in blazars remains incompletely understood. 
Electrons are likely energized through a combination of shock acceleration \cite[e.g.,][]{summerlin2012,marscher2014}, stochastic scattering \cite[e.g.,][]{katarzynski2006,lefa2011,asano2015,baring2017}, or electrostatic acceleration driven by magnetic reconnection \cite[e.g.,][]{giannios2009,giannios2013,petropoulou_2016,sironi2016}. 
Accretion disk instabilities around the central black hole have also been suggested as a possible acceleration mechanism \cite[e.g.,][]{sun2021,ha2025}.
Alongside these theoretical approaches, it is common to assume functional forms for the emitting electron distribution by modeling observed SEDs \cite[e.g.,][]{hayashida2012,dermer2014,yan2015,zheng2016,rahman2023}.
Some studies have developed self-consistent models including particle acceleration, escape, and energy losses by solving the Fokker--Planck equation \cite[e.g.,][]{lewis2016,lewis2018,lewis2019,muller2020}. 
It has particularly been shown in~\cite{lewis2018} that a one-zone leptonic model, where jet emission originates predominantly from a single emitting region (the jet blob), can successfully reproduce the observed blazar SEDs. 
In this scenario, nonthermal electrons accelerated within the jet blob are isotropic in the blob's co-moving frame and emit synchrotron radiation while also upscattering background photons via Compton scattering.

In 1959, S.~L.~Glashow had predicted that the cross section for electron antineutrino--electron scattering should be resonantly enhanced due to the production of on-shell $W^{-}$ bosons in the s-channel $\bar{\nu}_{e} e^{-} \rightarrow W^{-}$ \cite{glashow1960}. 
Today, this process is usually referred to as the Glashow resonance. 
To produce the resonance (with a $W^{-}$ boson mass of approximately $80.3 \, \mathrm{GeV}$) through scattering with electrons at rest, the incident (anti)neutrino must have an energy of about $6.3 \, \mathrm{PeV}$ \cite{berezinsky1977}. 
Such high-energy neutrinos could come from cosmic rays, but the corresponding relatively low flux on Earth apparently requires significant observation times even with large-scale detectors like IceCube. 
For instance, after over a decade of observations, IceCube has detected only a few PeV-scale neutrino events, with one candidate event occurring near the resonance energy at $6.05 \pm 0.72 \, \mathrm{PeV}$ \cite{aartsen2021}. 
Although many studies have explored the implications and potential signatures of detection of the Glashow resonance, independent experimental confirmations of its existence remain elusive \cite{aartsen2021,distefano2021}.

Recently, it has been proposed that, beyond the standard neutrino-induced channel, the Glashow resonance could be excited in electron--positron collisions, without a primary neutrino beam \cite{alikhanov2025}. 
This process involves high-energy electron--positron pairs producing on-shell $W^{\pm}$ bosons through the interaction $e^+ e^- \to W^\pm \rho(770)^{\mp}$. 
In addition to the latter resonant channel for $W^{\pm}$ bosons, high-energy electron--positron pairs can also undergo annihilation into $Z$ bosons, $e^{+} e^{-} \rightarrow Z$, when the center-of-mass energy approaches the $Z$ mass ($\sim 91 \, \mathrm{GeV}$). 
This process features a much larger cross section compared to the Glashow resonance and produces all three neutrino flavors through $Z$ decay. 
Considering Doppler boosting in blazar jets ($\delta_{D} \sim 70$) and redshift effects, the resulting neutrinos are expected to arrive at Earth primarily in the sub-TeV to TeV energy range. 
Among blazar subclasses, FSRQs provide particularly favorable environments to investigate this reaction due to their characteristically high electron and positron number densities within the jet. 
Leveraging electron energy distributions derived from SED modeling of FSRQs, we numerically estimate the reaction rates for both the $W^{\pm}$ and $Z$ channels and evaluate their possible contributions to the high-energy neutrino flux.
This synthesis of particle physics and blazar jet phenomenology provides a novel framework to probe electroweak interactions, including rare processes, in extreme astrophysical environments.

Building on these theoretical foundations, our study also incorporates electron number distributions derived from blazar SED modeling to investigate the role of the electron--positron collision channels within the jet environment. 
We further explore the astrophysical implications of such interactions, assessing their potential impact on the high-energy particle population and the resulting radiation spectrum observable on Earth. 
This connection between the particle physics proposal and blazar jet modeling offers a promising interdisciplinary avenue to investigate new particle interaction channels in extreme astrophysical settings.

\begin{figure}[t]
    \centering
    \includegraphics[width=1\linewidth]{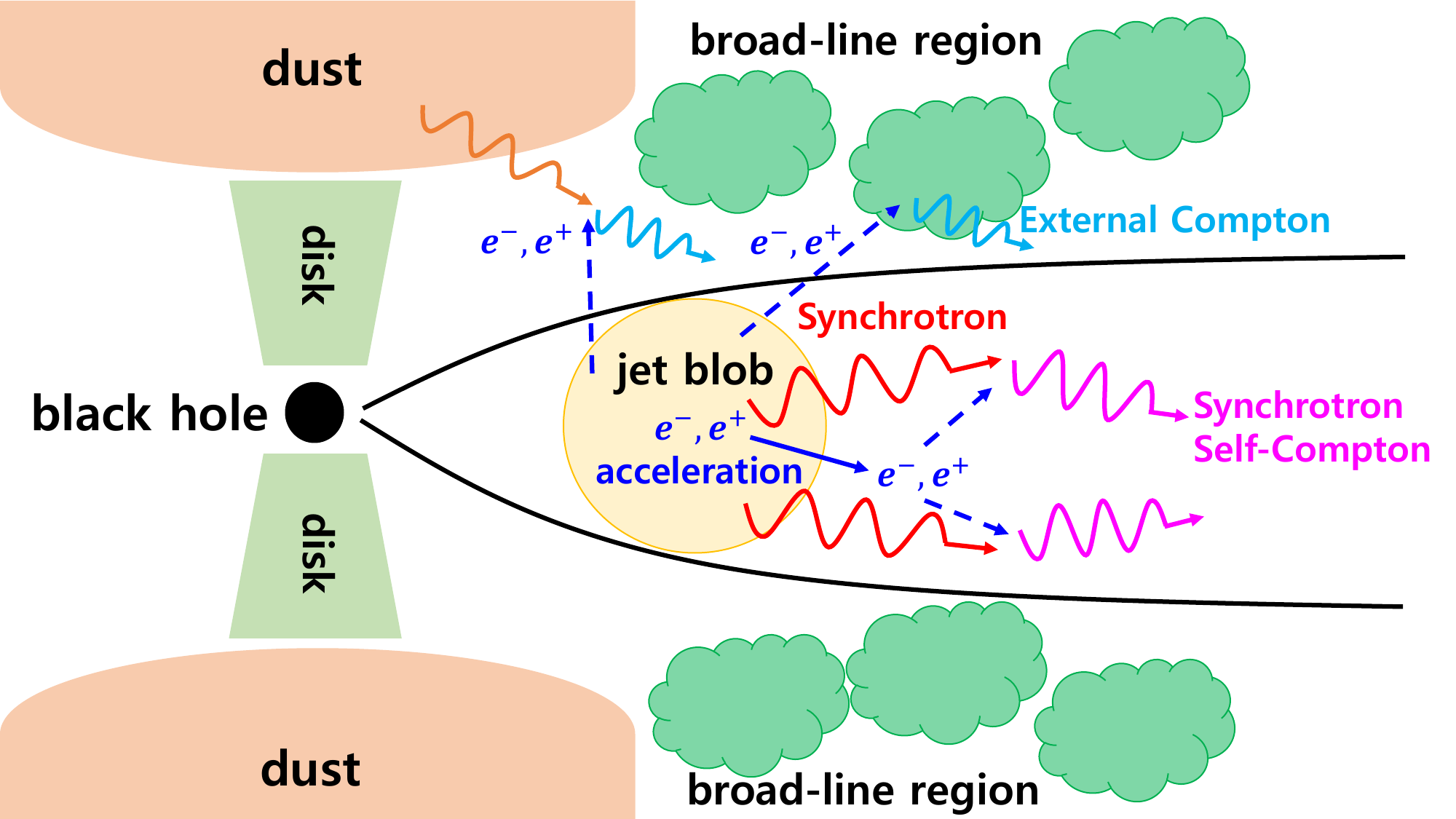}
    \caption{Schematic diagram of a one-zone leptonic model of a blazar jet. Particle acceleration primarily occurs within the jet blob, where accelerated electrons emit synchrotron radiation. These synchrotron photons serve as seed photons for Compton scattering, known as Synchrotron Self-Compton. The model also includes external Compton scattering involving photons from the dust torus and broad-line region.
    }
    \label{fig:f1}
\end{figure}

\section{Acceleration model in the jet blob}
\label{sec:s2}

\subsection{Model description}
\label{sec:s2.1}

The blazar jet is possibly launched by a rapidly spinning black hole threaded with magnetic fields anchored in the surrounding accretion disk \cite{blandford1977}. 
The jet plasma moves toward the observer with a bulk Lorentz factor $\Gamma = {(1 - \beta^2)}^{-1/2}$.
Because the plasma travels relativistically at a small angle $\theta$ to the line of sight, the observed emission is Doppler-boosted by a factor $\delta_D = [\Gamma(1-\beta~{\rm cos}\theta)]^{-1}$.
For simplicity, we adopt the common approximation $\Gamma = \delta_{D}$, which is valid under typical blazar conditions, where the jet is closely aligned with the observer's line of sight. In such cases, with small viewing angles ($\theta \approx \Gamma^{-1}$) and highly relativistic speeds ($\beta \approx 1$), the Doppler factor approximates to $\delta_{D} \approx \Gamma$. 
In the one-zone leptonic model illustrated in Figure~\ref{fig:f1}, the primary emitting region is modeled as a compact, homogeneous blob. 
This co-moving blob is widely regarded as the dominant source of radio emission, where accelerated electrons produce synchrotron radiation \cite[e.g.,][and references therein]{lewis2019}.

We solve the Fokker--Planck equation to examine electron acceleration in the jet blob. 
Both diffusive shock acceleration and stochastic acceleration due to turbulence are considered as mechanisms responsible for energizing electrons. 
In addition, the model accounts for energy losses from adiabatic expansion and radiative processes, particle escape via Bohm diffusion, and continuous injection of monoenergetic electrons. 
The nonthermal electron population in the blob is modeled in the frame co-moving with the blob. 
The governing Fokker--Planck equation takes the following form:
\begin{equation}
\frac{\partial N_e}{\partial t} 
= \frac{\partial^2}{\partial \gamma^2} \left( D_0 \, \gamma^2 \, N_e \right) 
- \frac{\partial}{\partial \gamma} \left( \left<\frac{d \gamma}{d t} \right>  N_e \right) 
- \frac{\gamma N_e D_0}{\tau} 
+ \dot{N}_{\rm inj} \, \delta(\gamma - \gamma_{\rm inj}),
\end{equation}
where $N_e(\gamma,t)$ is the electron number distribution, and $\gamma \equiv {E}/{m_e c^2}$ is the electron Lorentz factor, with $m_e$ and $c$ denoting the electron mass and the speed of light, respectively. 
The parameters $\dot{N}_{\rm inj}$ and $\gamma_{\rm inj}$ represent the injection rate and the Lorentz factor of the monoenergetic electrons, respectively.

In the steady-state, the Fokker--Planck equation is reduced to
\begin{equation}
\frac{\partial^2}{\partial \gamma^2} \left( D_0 \, \gamma^2 \, N_e \right) 
- \frac{\partial}{\partial \gamma} \left( \left<\frac{d \gamma}{d t} \right> N_e \right) 
- \frac{\gamma N_e D_0}{\tau} 
+ \dot{N}_{\rm inj} \, \delta(\gamma - \gamma_{\rm inj}) = 0.
\end{equation}
The first term represents spectral broadening due to energy diffusion, characterized by the diffusion coefficient $D_0$. 
The second term accounts for energy gains from stochastic acceleration in turbulence and diffusive shock acceleration at shocks, as well as energy losses from adiabatic expansion and radiative processes (i.e., synchrotron radiation and inverse Compton scattering). 
The third term describes the escape of electrons from the jet blob via Bohm diffusion. 
The steady-state electron distribution function is obtained by solving this equation using the analytical approach and parameter sets provided in~\cite{lewis2019}.

The coefficient $<{d\gamma}/{dt}>$ includes acceleration processes and losses, and is given by
\begin{equation}
\left<\frac{d\gamma}{dt}\right> = \dot{\gamma}_{\rm sto} + \dot{\gamma}_{\rm ad+sh} - \dot{\gamma}_{\rm syn} - \dot{\gamma}_{C}.
\end{equation}
Here, $\dot{\gamma}_{\rm sto}$ represents the stochastic acceleration mediated by turbulence and depends on the diffusion coefficient $D_0$:
\begin{equation}
\dot{\gamma}_{\rm sto} = 4 D_0 \, \gamma.
\end{equation}
$\dot{\gamma}_{\rm ad+sh}$ accounts for both diffusive shock acceleration and adiabatic losses:
\begin{equation}
\dot{\gamma}_{\rm ad+sh} = a D_0 \, \gamma,
\end{equation}
where $a$ is a dimensionless parameter characterizing the efficiency of diffusive shock acceleration and adiabatic losses. 
For $a>0$, diffusive shock acceleration dominates over adiabatic losses, while for $a<0$, adiabatic losses dominate.
The synchrotron loss rate $\dot{\gamma}_{\rm syn}$ reads
\begin{equation}
\dot{\gamma}_{\rm syn} = D_0 b_{\rm syn} \gamma^2 = \frac{\sigma_T B^2}{6 \pi m_e c} \gamma^2,
\end{equation}
where $\sigma_T = 6.65 \times 10^{-25} \, {\rm cm^2}$ is the Thomson cross section, and $B$ is the strength of the tangled, homogeneous magnetic field.  
The Compton cooling rate is included in $<{d\gamma}/{dt}>$. 
The Compton scattering refers to the scattering of external seed photons that impinge on the jet and collide with electrons in the blob. 
As external media, the dust torus and the broad-line region (BLR) are considered. The Compton cooling rate $\dot{\gamma}_C$ is given by
\begin{equation}
\dot{\gamma}_C = \sum_{j=1}^{J} D_0 \, \gamma^2 \, b_C^{(j)} \, H(\gamma_{\rm ph}^{(j)}) 
= \sum_{j=1}^{J} \frac{4 \sigma_T \Gamma^2 u_{\rm ph}^{(j)}}{3 m_e c} \gamma^2 \, H(\gamma_{\rm ph}^{(j)}),
\end{equation}
where $b_C^{(j)}$ is a dimensionless constant related to Compton cooling for the different external radiation fields $j$, with energy density $u_{\rm ph}^{(j)}$. 
As modeled in previous works~\cite{lewis2019,lewis2018}, we include a dust torus ($b_C^{(\rm dust)}$) and a broad line region ($b_C^{(\rm BLR)}$).  
The function $H(\gamma_{\rm ph}^{(j)})$ represents the mitigation of energy losses with the full Compton cross-section, including the Klein--Nishina effect \cite{boettcher1997}, and depends on the incident photon energy $\gamma_{\rm ph}^{(j)}$. 
It is defined by
\begin{equation}
H(y) \equiv \frac{9}{32} \frac{1}{y^2} G(y),
\end{equation}
where $G(y)$ has the form
\begin{align}
G(y) &= \frac{8}{3} \frac{y (1+5y)}{(1+4y)^2} - \frac{4y}{1+4y} \left( \frac{2}{3} + \frac{1}{2y} + \frac{1}{8y^2} \right) \nonumber \\
&\quad + \ln(1+4y) \left[ 1 + \frac{3}{y} + \frac{3}{4y^2} + \frac{\ln(1+4y)}{2y} - \frac{\ln(4y)}{y} \right] 
+ \frac{1}{y} \sum_{n=1}^{\infty} \frac{(1+4y)^{-n}}{n^2} - \frac{5}{2y} - \frac{\pi^2}{6y} - 2.
\end{align}

Along with the acceleration processes and losses, the energy-dependent particle escape term is included in the steady-state Fokker--Planck equation. 
The dimensionless escape parameter $\tau$, based on Bohm diffusion, is given by
\begin{equation}
\tau = \frac{R_b^2 q B D_0}{m_e c^3},
\end{equation}
with $q$ being the fundamental charge.  
Additionally, the particle injection term is defined by the particle injection rate
\begin{equation}
\dot{N}_{\rm inj} = \frac{L_{e,{\rm inj}}}{\gamma_{\rm inj} m_e c^2},
\end{equation}
where $L_{e,{\rm inj}}$ is the electron injection luminosity and $\gamma_{\rm inj}$ is the injection Lorentz factor. 
While both $L_{e,{\rm inj}}$ and $\gamma_{\rm inj}$ are free parameters, we adopt values that well reproduce the observed SED.

The normalization of the electron distribution function can be inferred from multi-wavelength observations, spanning from radio to gamma-ray. 
Electrons accelerated in the jet blob emit synchrotron radiation, synchrotron-self Compton (SSC), and external Compton (EC) scattering, which together contribute to the spectral energy distribution (SED) across the radio to gamma-ray wavelengths. 
Based on SED modeling from radio to gamma-ray \cite[e.g.,][]{yan2015,lewis2018,rahman2023}, the electron jet power can be estimated as
\begin{equation}
P_e = 2 \pi R_b^2 \, \beta c \, \Gamma^2 \, u_e.
\end{equation}
Here, $u_e$ is the electron energy density, given by
\begin{equation}
u_e = \frac{m_e c^2}{V_b} \int d\gamma \, \gamma N_e(\gamma),
\end{equation}
where $V_b = {4\pi}R_b^3/{3}$ is the volume of the jet blob with radius $R_b$.

In this work the stochastic acceleration coefficient $D_{\rm sto} \propto \gamma^2$ corresponds to the so called hard sphere scattering regime in which the scattering mean free path is independent of particle energy.
Such a prescription is commonly adopted in one zone models of blazar jets and is motivated by numerical studies of strong turbulence with $\delta B/B \gtrsim 1$ \cite[e.g.,][]{asano2015, baring2017}.
In contrast weak turbulence is expected to yield an energy dependent diffusion coefficient $D_{\rm sto} \propto \gamma^q$ with $q = 1/3$ for Kolmogorov type spectra.
To assess the sensitivity to this assumption we have verified that adopting a Kolmogorov like scaling leads to a modest steepening of the steady state electron distribution and results in changes of less than a factor of two in the predicted neutrino flux.
Our main conclusions are therefore robust against the specific choice of the stochastic acceleration regime.

We also note that recent studies of particle transport in strongly magnetized and intermittent turbulence have demonstrated that spatial diffusion can exhibit a non trivial energy dependence driven by field line reversals and regions of large magnetic field curvature rather than following a simple Bohm like scaling \cite[e.g.,][]{kempski2023}.
In the present work, however, our treatment of particle transport is phenomenological and anchored to one zone spectral energy distribution modeling where the electron/positron distribution is constrained by observations.
Since the electroweak interaction rates considered here depend primarily on the total pair density in the emitting region and on the fraction of particles near the resonant energy the detailed microphysics of spatial diffusion does not affect our main results at the order-of-magnitude level.

\begin{table}[htbp]
\centering
\begin{tabular}{c|c|c}
\hline
Parameter (Unit) & Model 1 & Model 2\\
\hline
$a$ & $-2.0$ & $-0.5$\\
$D_0~({\rm s}^{-1})$ & $9.0 \times10^{-6}$ & $1.5 \times10^{-5}$\\
$\gamma_{\rm inj}$ & $1.01$ & $2.01$ \\
$L_{e,inj}~({\rm erg~s^{-1}})$ & $8.5 \times10^{30}$ & $5.4 \times10^{30}$ \\
$R_{b}~({\rm cm})$ & $2.4 \times10^{15}$ & $1.2 \times10^{15}$ \\
$u_{ph}^{({\rm dust})}~({\rm erg~cm^{-3}})$ & $1.6 \times10^{-4}$ & $1.6 \times10^{-4}$ \\
$u_{ph}^{({\rm BLR})}~({\rm erg~cm^{-3}})$ & $4.2 \times10^{-4}$ & $1.1 \times10^{-3}$ \\
$B~({\rm G})$ & $0.21$ & $0.30$ \\
$\delta_D,~\Gamma$ & $49$ & $70$ \\
$P_e~({\rm erg~s^{-1}})$ & $9.0 \times10^{45}$ & $4.3 \times10^{45}$ \\
$u_e~({\rm erg~cm^{-3}})$ & $3.45$ & $3.23$ \\
\hline
\end{tabular}
\caption{Parameters of 3C 279 during the flaring state of December 20, 2013, used to calculate the electron energy distribution, adopted from \cite{lewis2019}.\label{tab:t1}}
\end{table}

\begin{figure}[t]
    \centering
    \includegraphics[width=0.7\linewidth]{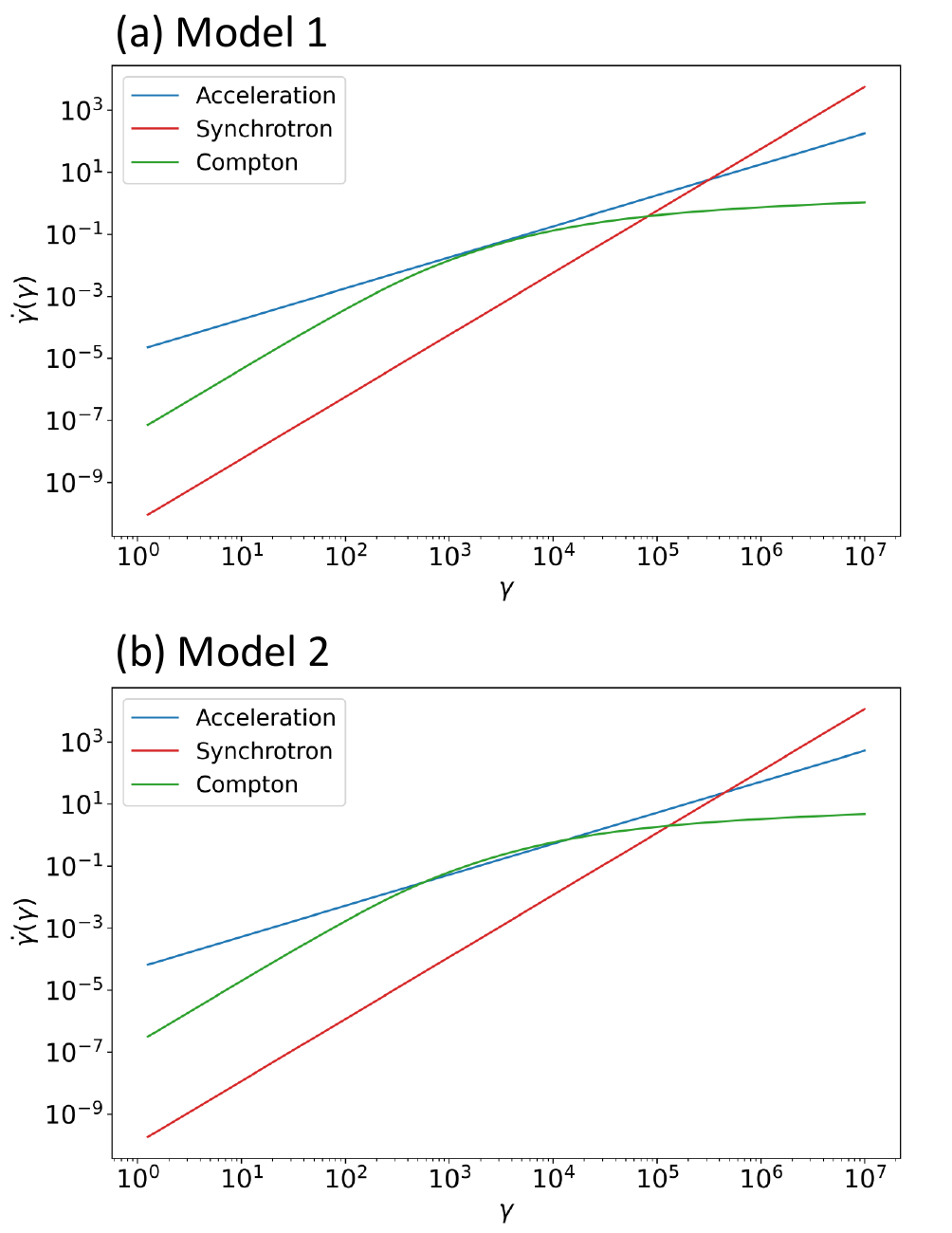}
    \caption{Rates of acceleration and energy losses via synchrotron and Compton scattering as functions of the electron Lorentz factor. The acceleration term includes diffusive shock acceleration, stochastic acceleration by turbulence, and adiabatic losses. The Compton cooling rate accounts for external Compton scattering involving seed photons from the dust torus and broad-line region.
    }
    \label{fig:f2}
\end{figure}

\begin{figure}[t]
    \centering
    \includegraphics[width=0.7\linewidth]{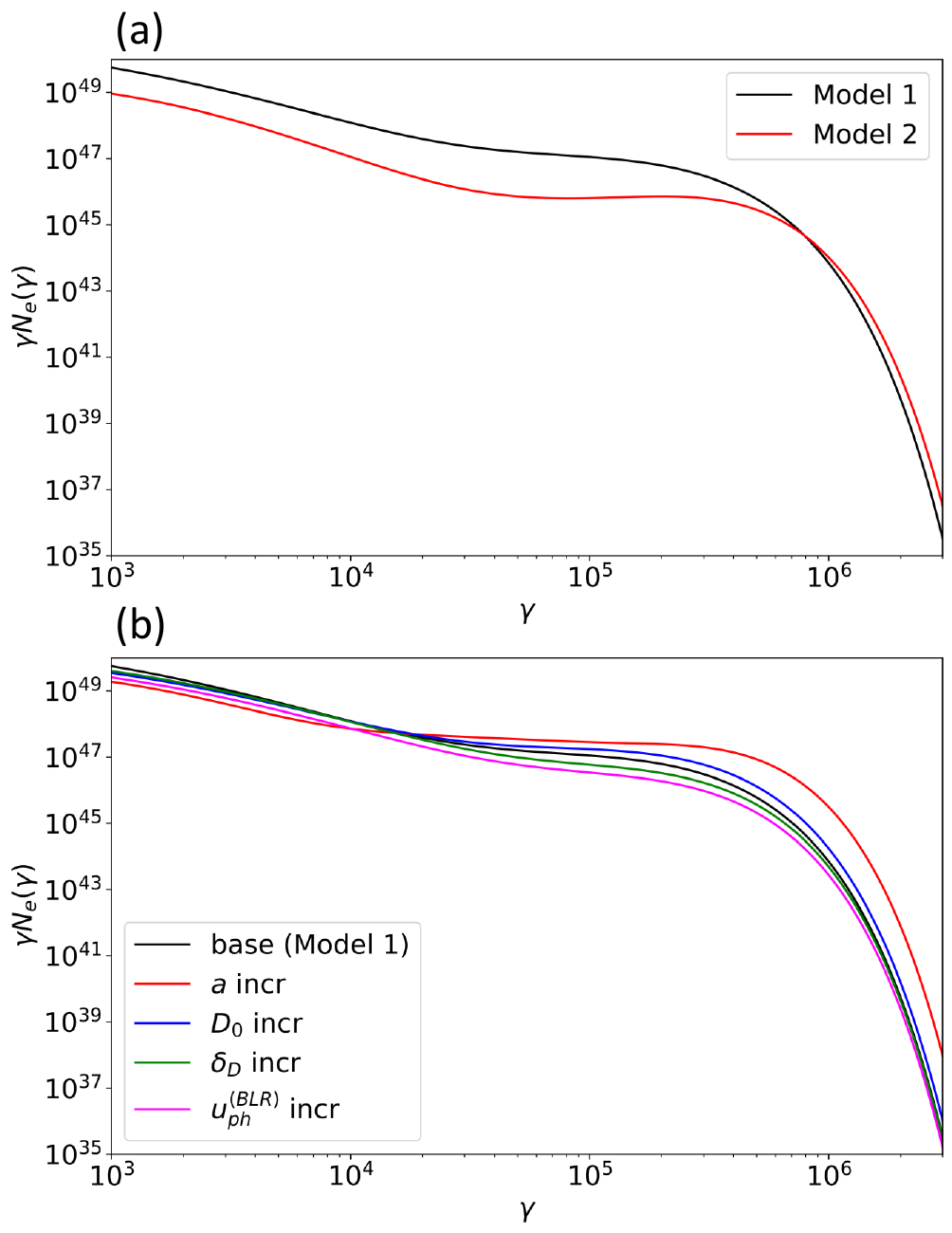}
    \caption{ (a) Steady-state electron energy distribution as a function of the Lorentz factor in the jet blob. (b) Based on the baseline model referred to as Model~1, we vary the acceleration parameter $a$, the diffusion coefficient $D_0$, the Doppler factor $\delta_D$, and the photon energy density in the BLR regions, $u_{ph}^{(\rm BLR)}$.}
    
    \label{fig:f3}
\end{figure}

\subsection{Modeling electron energy distribution in 3C 279}
\label{sec:s2.2}

We specifically model the FSRQ 3C 279 during its flaring state on December 20, 2013, which has been extensively studied in previous works \cite[e.g.,][]{asano2015,hayashida2015,paliya2016,lewis2016,lewis2019}. 
Table~\ref{tab:t1} summarizes the parameters used to fit the observed SED of 3C 279 during this flaring state, as obtained in~\cite{lewis2019}.
The SED modeling is based on synchrotron, SSC, and external Compton emissions from accelerated electrons within the jet blob. 
Since the observed spectrum covers a limited frequency range, multiple combinations of parameters can produce results consistent with the observations.
Consequently, there is an intrinsic level of uncertainty in key parameters such as the acceleration parameter $a$, the turbulent diffusion coefficient $D_0$, the Doppler factor $\delta_D$, and the photon energy densities in the dust and BLR regions. 
The parameters $a$ and $D_0$ govern acceleration, energy losses, and diffusion within the jet blob, while the Doppler factor and photon energy densities determine radiative cooling \cite[e.g.,][]{lewis2016,lewis2018}.
Since the electron--positron collision processes considered in this work occur at energies $\sim 100$~GeV, the associated uncertainties in these physical parameters can significantly affect the predicted electron distributions and, subsequently, the corresponding reaction rates in the blazar jet.

Figure~\ref{fig:f2} presents the characteristic rates for acceleration and cooling in the jet blob of 3C 279. 
In the low-energy range ($\gamma \lesssim 10^2$), acceleration dominates over radiative cooling via synchrotron and Compton scattering. 
In the intermediate energy range ($\gamma \sim 10^3 - 10^4$), while cooling via Compton scattering becomes significant, acceleration still plays a crucial role in generating high-energy electrons. 
The Compton loss rate changes at higher energies (around $\gamma \sim 10^5$) due to the Klein--Nishina effect, causing the electron distribution to become harder in the energy range around $\gamma \sim 10^5$. 
For energies $\gamma \gtrsim 10^6$, synchrotron cooling dominates over other processes, and the maximum electron energy is likely determined by synchrotron cooling.  

Figure~\ref{fig:f3}-(a) shows the steady-state solution of the Fokker--Planck equation for the expected electron energy distribution in 3C 279.
As shown in Figure~\ref{fig:f2}, since acceleration dominates over the cooling rate associated with Compton scattering, a flat electron bump is formed in the energy range $\gamma \sim 10^4 - 10^5$. 
Specifically, in the model parameter, a negative value of $a$ indicates that shock acceleration is inefficient compared to adiabatic losses.
In this context, the balance between stochastic acceleration mediated by turbulence and adiabatic losses results in a flat spectral slope.
The exponential cutoff observed at higher energies is due to synchrotron cooling for $\gamma \gtrsim 10^6$.

In Figure~\ref{fig:f3}-(b), we further examine how the electron energy distribution is influenced by key parameters of the jet blob, including the acceleration parameter $a$, the diffusion coefficient $D_0$, the Doppler factor $\delta_D$, and the photon energy density in the BLR regions $u_{ph}^{(\rm BLR)}$. 
Increasing the acceleration parameter $a$ flattens the distribution due to enhanced diffusive shock acceleration and/or the suppression of adiabatic cooling. 
The spectral slope becomes harder as $D_0$ increases, owing to the additional energy gain from turbulent stochastic acceleration. 
Cooling effects depend on the Doppler factor and photon energy density in the BLR, which govern Compton scattering. 
Specifically, the spectral slope becomes softer as either $\delta_D$ or $u_{ph}^{(\rm BLR)}$ increase due to stronger radiative cooling. 
According to current modeling studies of blazar jet SEDs, such uncertainties in the electron energy distribution are expected, and they become more pronounced at higher energies with $\gamma \gtrsim 10^5$.

\section{Electron--positron interactions and neutrino emission in blazar jets}
\label{sec:s3}

In this section, we present the neutrino emission in blazar jets associated with electron--positron interactions.
Before introducing the individual interaction channels we clarify the formulation adopted throughout this section.
The electroweak bosons $W^{\pm}$ and $Z$ produced in electron--positron collisions are extremely short lived and decay on timescales that are many orders of magnitude shorter than any relevant dynamical timescale of the jet emitting region.
Consequently, treating the interaction in terms of resonant on shell $W^{\pm}$ or $Z$ production followed by prompt decay is fully equivalent to directly computing neutrino production from electron--positron annihilation into neutrino final states.
We implicitly account for the relevant branching ratios through the decay of the electroweak bosons such that the effective neutrino yield is identical to that obtained from a direct $e^+ e^- \to \nu \bar{\nu}$ treatment.
We nevertheless formulate the calculation in terms of explicit $W^{\pm}$ and $Z$ production in order to highlight the role of electroweak resonances their characteristic energy scales and their branching into neutrino final states.
This choice provides a transparent physical interpretation of the interaction channels and does not affect the resulting neutrino spectra or flux normalization.

\subsection{Resonant W boson production via electron--positron interactions in the jet from 3C 279}
\label{sec:s3.1}

We primarily examine the possibility of excitation of the Glashow resonance via electron--positron collisions in the jet blob of AGNs. 
This idea is inspired by a recent theoretical study demonstrating that such a resonance could occur in the absence of a primary neutrino beam \cite{alikhanov2025}. 
In this scenario, high-energy electron--positron collisions can produce on-shell $W^{\pm}$ bosons through the $s$-channel subprocess in $e^+ e^- \rightarrow W^{\pm} \, \rho(770)^{\mp}$.
While this mechanism has been considered in the context of high-luminosity electron--positron colliders, we evaluate its feasibility in astrophysical environments such as blazars, which are regarded as efficient accelerators of electrons and positrons. 
It is interesting to note that the proposed process allows one to probe the CP conjugate of the Glashow resonance, providing thus an important test of the neutrino sector of the Standard Model in the resonance region.

While particles are accelerated within the compact jet blob region, we consider that the dominant electron--positron interactions occur within the blob itself, where the lepton density is the highest. According to numerical works on acceleration of pair plasma in relativistic flows \cite[e.g.,][]{sironi2013,plotnikov2018}, the number of accelerated positrons is expected to be roughly equal to the number of accelerated electrons. Using the number density of $\sim 50 \, \mathrm{GeV}$ electrons and positrons within the blob, the volume-averaged number density of these particles is estimated as
\begin{equation}
\bar{n}_{e^-}^{50 \, \mathrm{GeV}} \sim \bar{n}_{e^+}^{50 \, \mathrm{GeV}}  
\sim \frac{N_e(\gamma \approx 9.8 \times 10^4) \, \Delta \gamma}{V_b} \,  
\sim 7.2 \, \mathrm{cm}^{-3}.
\end{equation}
Adopting the cross section of the electron--positron mediated Glashow resonance $\sigma_{e^- e^+}^{W^{\pm}} \sim 10^{-42} \, \mathrm{cm}^2$ \cite{alikhanov2025}, the total reaction rate in the blazar jet reads
\begin{equation}
\Gamma_{e^- e^+}^{W^{\pm}} \sim n_{e^-}^{50 \, \mathrm{GeV}} \, n_{e^+}^{50 \, \mathrm{GeV}} 
\, \sigma_{e^- e^+}^{W^{\pm}} \, c \, V_b
\sim 2.9 \times 10^{16} \, \mathrm{s}^{-1}.
\end{equation}
We note that the diffusion timescale for electrons to escape the blob is much longer than the typical timescale for electron--electron and electron--positron interactions within the blob. 
Using the parameters $R_b = 2.4 \times 10^{15}~\mathrm{cm}$, $B = 0.21~\mathrm{G}$, $D_0 = 9.0 \times 10^{-6}$, and $\gamma = 9.8 \times 10^4$, the diffusive escape timescale can be expressed in terms of the dimensionless escape parameter $\tau$ as
\begin{equation}
t_{\rm esc} = \frac{\tau}{\gamma D_0} = \frac{R_b^2eB}{\gamma m_e c^3} \sim 2.4 \times 10^{11}~{\rm s},
\end{equation}
which is many orders of magnitude longer than both the blob dynamical timescale, $t_{\rm dyn} \sim R_b/c \sim 8 \times 10^4~{\rm s}$.
We note that $\Gamma_{e^- e^+}^{W^\pm}$ represents the total interaction rate integrated over the blob volume, corresponding to a characteristic global interaction timescale $t_{\rm e^-e^+}^{\rm W^{\pm}} \sim 1/\Gamma_{e^- e^+}^{W^\pm} \sim 3 \times 10^{-17}~\mathrm{s}$.
This ensures that essentially all interactions occur before electrons can escape, justifying the in-blob treatment of electroweak processes.

Multiplying by the characteristic energy of $100 \, \mathrm{GeV}$, the energy budget associated with $W^{\pm}$ boson production is estimated as
\begin{equation}
L_{e^- e^+}^{W^{\pm}} \sim 100 \, \mathrm{GeV} \times \Gamma_{e^- e^+}^{W^{\pm}} 
\sim 4.6 \times 10^{15} \, \mathrm{erg \, s^{-1}},
\end{equation}
which is many orders of magnitude smaller than the jet power (i.e., an order of $10^{45}~{\rm erg~s^{-1}}$).
While the total number of electron--positron collisions in a blazar jet may reach $\sim 10^{16} \, \mathrm{s}^{-1}$, corresponding to $\gtrsim 10^{23}$ events $\mathrm{yr}^{-1}$, future leptonic colliders like CEPC \cite{cheng2022} and FCC-ee \cite{agapov2022} are expected to observe an order of 10 events annually under controlled conditions~\cite{alikhanov2025}.
This comparison highlights the complementarity between high-energy astrophysical environments, which provide vast natural laboratories for rare electroweak interactions, and precision collider experiments, which offer clean, controllable settings to verify the microphysics of such processes.

\begin{figure}[t]
    \centering
    \includegraphics[width=1\linewidth]{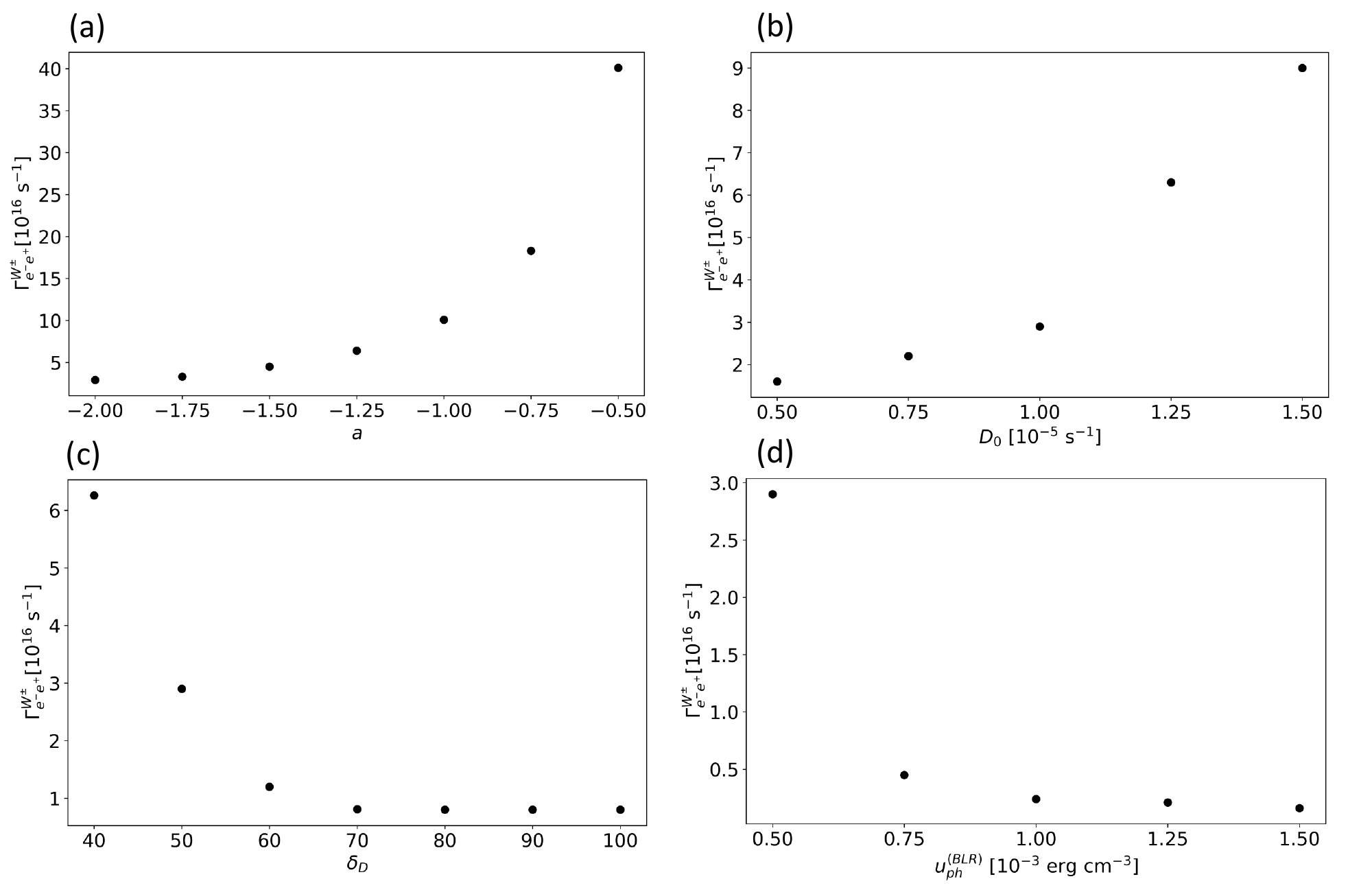}
    \caption{ Dependences of the reaction rate for $e^+ e^- \rightarrow W^{\pm} \rho(770)^{\mp}$ on: (a) the acceleration parameter $a$; (b) the diffusion coefficient $D_0$; (c) the Doppler factor $\delta_D$; (d) the photon energy density in the BLR regions $u_{ph}^{(\rm BLR)}$. For all panels, each parameter acts as free from the parameter set of Model~1.}
    \label{fig:f4}
\end{figure}

We further examine the significance of the parameters that represent acceleration and energy loss within the jet blob. 
Four key parameters are considered, each of which can modify the electron energy distribution: 
(a) the acceleration parameter $a$, which characterizes the balance between shock acceleration and adiabatic losses; 
(b) the diffusion coefficient $D_0$, associated with turbulent stochastic acceleration; 
(c) the Doppler factor $\delta_D$, which determines both the electron jet power and radiative cooling via synchrotron and Compton scattering; 
(d) the photon energy density in the BLR regions, $u_{ph}^{(\rm BLR)}$, which controls radiative cooling. 
Based on the well-fitted cases listed as Model~1 and Model~2, plausible values of $a = [-2.0, -0.5]$, $D_0 = [0.5, 1.5] \times 10^{-5}~{\rm s^{-1}}$, $\delta_D = [40, 100]$, and $u_{ph}^{(\rm BLR)} = [0.5, 1.5] \times 10^{-3}~{\rm erg~cm^{-3}}$ are adopted.  

Figure~\ref{fig:f4} shows the resulting reaction rate for $e^+ e^- \rightarrow W^{\pm} \, \rho(770)^{\mp}$ as a function of $a$, $D_0$, $\delta_D$, and $u_{ph}^{(\rm BLR)}$. 
Except for the main variable shown in each panel of Figure~\ref{fig:f4}, all other parameters used for the electron energy distribution are adopted from Model~1. 
The main results are consistent with the parameter dependence of the electron energy distribution shown in Figure~\ref{fig:f3}.  
As expected, a larger $a$ enhances the efficiency of shock acceleration, leading to an increase in the electron number density at $100 \, \mathrm{GeV}$ and a corresponding increase in the reaction rate. 
Similarly, enhancing turbulent stochastic acceleration (i.e., increasing $D_0$) also increases the reaction rate. 
The cooling effects are examined in panels (c) and (d). It is found that the reaction rate decreases as $\delta_D$ increases. 
In the range $\delta_D \lesssim 70$, radiative cooling dominates over the enhancement of jet power, whereas the two effects are approximately balanced for $\delta_D \gtrsim 70$. 
Additionally, the reaction rate decreases as $u_{ph}^{(\rm BLR)}$ increases due to enhanced radiative cooling via Compton scattering.  
Although the rate varies by up to an order of magnitude across the examined parameter ranges, the overall impact remains subdominant compared to the total particle injection energy rate inside the jet blob. 
Therefore, the inclusion of this exotic channel does not qualitatively alter the general energy budget or the modeling framework for blazar SEDs.

To evaluate the detectability of such events on Earth, we estimate the corresponding observable diffuse neutrino flux from 3C~279 ($z \sim 0.536$), using the luminosity distance ($d_L \sim 3.1 \, \mathrm{Gpc}$).
The resulting value is
\begin{equation}
F_{e^- e^+}^{W^{\pm}, \rm obs} \sim 
\frac{\Gamma_{e^- e^+}^{W^{\pm}}}{4 \pi d_L^2 (1+z)} 
\sim 1.7 \times 10^{-40} \, \mathrm{cm^{-2} \, s^{-1}},
\end{equation}
which lies many orders of magnitude below the sensitivity thresholds of current neutrino telescopes such as IceCube \cite[e.g.,][]{aartsen2017}, KM3NET \cite[e.g.,][]{aiello2024}, and Baikal-GVD \cite[e.g.,][]{aynutdinov2023}.
We have assumed that each reaction eventually produces a neutrino. 
This is justified since the $W^{\pm}$ boson predominantly decays either directly into a lepton pair (as $\nu_e e^+$) or into hadrons, which would subsequently produce neutrinos (through meson decays, for example).
Note that, in these decays, neutrinos of all three flavors will appear.  
Although the Glashow resonance via electron--positron collisions is theoretically possible in astrophysical environments, the extremely low flux anticipated from AGNs renders direct detection virtually unfeasible with existing or near-future instrumentation. 
Nevertheless, the existence of this interaction channel in high-density astrophysical plasmas remains a theoretically compelling prospect. 
We have used the conservative value of the cross section. 
The latter can be modified by contributions beyond the Standard Model as well as by the momentum transfer-dependent couplings of the mesons to the leptonic current \cite{alikhanov2025}. 
It may inspire future observational strategies or reinterpretations of high-energy astrophysical phenomena, particularly under more extreme or localized conditions.

\begin{figure}[t]
    \centering
    \includegraphics[width=1\linewidth]{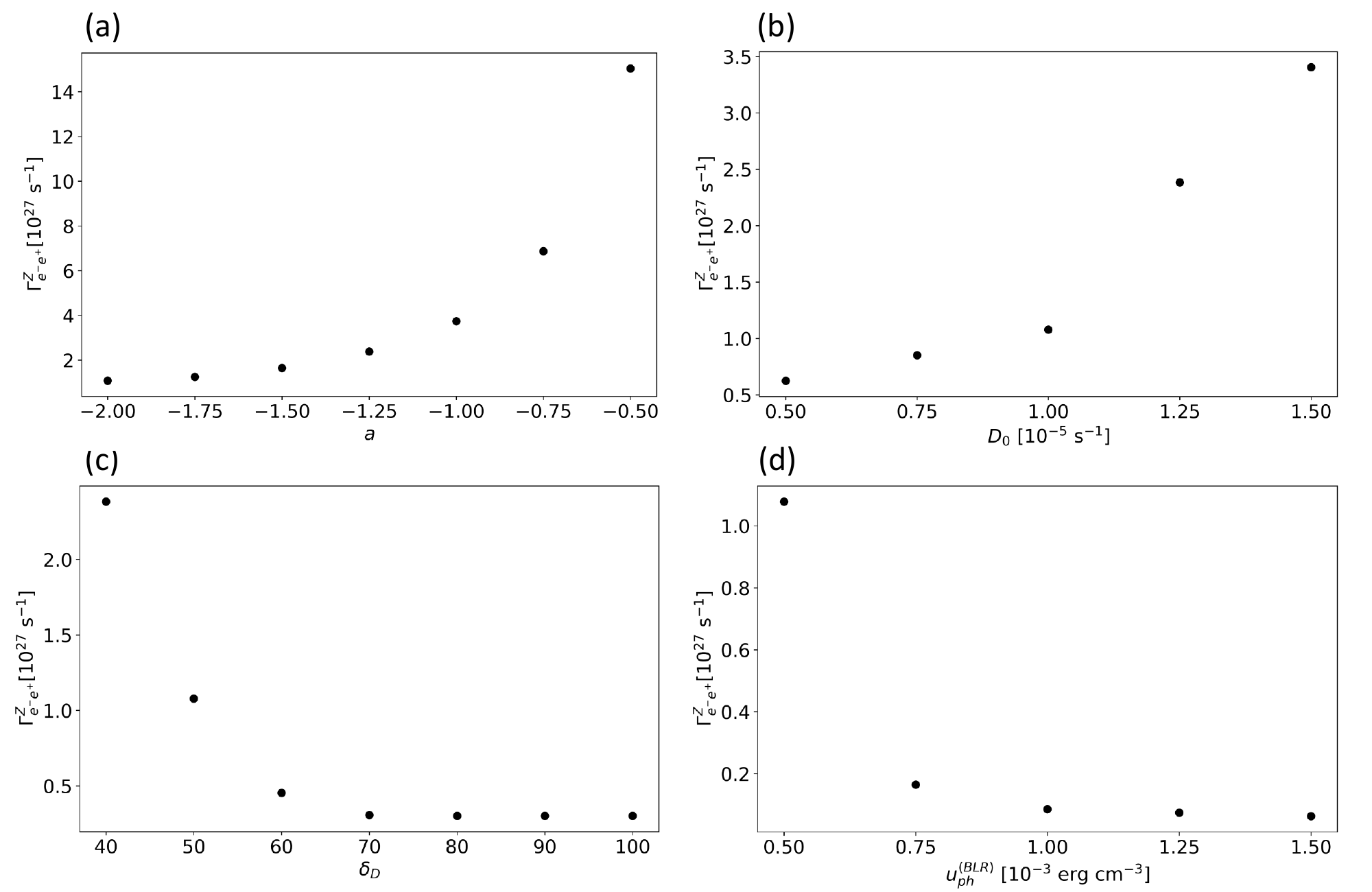}
    \caption{ Dependences of the reaction rate for $e^+ e^- \rightarrow Z$ on: (a) the acceleration parameter $a$; (b)  the diffusion coefficient $D_0$; (c) the Doppler factor $\delta_D$; (d) the photon energy density in the BLR regions $u_{ph}^{(\rm BLR)}$. For all panels, each parameter  acts as free from the parameter set of Model~1.}
    
    \label{fig:f5}
\end{figure}

\subsection{Resonant Z boson production via electron--positron interactions in the jet from 3C 279}
\label{sec:s3.2}

In addition to the $W$ boson production, high-energy electron--positron pairs in the jet can also undergo resonant annihilation into $Z$ bosons when the center-of-mass energy approaches the $Z$ boson mass ($m_Z \sim 91 \, \mathrm{GeV}$). 
This is not far from the $W$ boson pole considered above. 
The corresponding process, $e^+  e^- \rightarrow Z$, features a peak cross section that is approximately ten orders of magnitude larger than that of the resonant channel described in Section~\ref{sec:s3.1}, $\sigma_{e^- e^+}^{Z} \sim 4 \times 10^{-32} \, {\rm cm^2}$ \cite{LEP_SLD2006}.
Adopting the same number density of $\sim 50 \, \mathrm{GeV}$ electrons and positrons in the jet as derived in Section~\ref{sec:s3.1}, the mean particle density relevant for this channel remains
\begin{equation}
n_{e^-}^{50 \, \mathrm{GeV}} \sim n_{e^+}^{50 \, \mathrm{GeV}} \sim 7.2 \, \mathrm{cm^{-3}}.
\end{equation}
The total reaction rate is then
\begin{equation}
\Gamma_{e^- e^+}^{Z} \sim n_{e^-}^{50 \, \mathrm{GeV}} \, n_{e^+}^{50 \, \mathrm{GeV}} \, \sigma_{e^- e^+}^{Z} \, c \, V_b \sim 1.1 \times 10^{27} \, \mathrm{s^{-1}}.
\end{equation}
Multiplying by the characteristic energy of $100 \, \mathrm{GeV}$, the energy budget associated with $Z$ boson production is estimated as
\begin{equation}
L_{e^- e^+}^{Z} \sim 100 \, \mathrm{GeV} \times \Gamma_{e^- e^+}^{Z} \sim 1.8 \times 10^{26} \, \mathrm{erg \, s^{-1}}.
\end{equation}

To evaluate the detectability of such events on Earth, we compute the observable neutrino flux from 3C~279 ($z \sim 0.536$), using the luminosity distance ($d_L \sim 3.1 \, \mathrm{Gpc}$). The resulting flux is
\begin{equation}
F_{e^- e^+}^{Z, \rm obs} \sim \frac{\Gamma_{e^- e^+}^{Z}}{4 \pi d_L^2 (1+z)} \sim 6.8 \times 10^{-30} \, \mathrm{cm^{-2} \, s^{-1}}.
\end{equation}
This value, while still below the sensitivity of current neutrino telescopes such as IceCube \cite[e.g.,][]{aartsen2017}, KM3NeT \cite[e.g.,][]{aiello2024}, or Baikal-GVD \cite[e.g.,][]{aynutdinov2023}, is dramatically enhanced compared to the resonance channel $e^+ e^- \rightarrow W^{\pm} \, \rho(770)^{\pm}$. 
Importantly, the $Z$ boson decays also produce all three neutrino flavors, implying a broader range of potential astrophysical signatures. 
Again, due to the dominance of the leptonic and hadronic decay channels, we assume that each $Z$ boson eventually gives a neutrino. 
Although the predicted flux remains too small for direct detection, the vastly larger cross section of the $Z$ channel suggests that resonant annihilation into $Z$ bosons could play a non-negligible role in shaping the secondary particle environment in dense pair plasmas.
As shown in Figure~\ref{fig:f5}, we examine the parameter dependence of $Z$ boson production on acceleration, diffusion, and cooling processes. 
While the same trends discussed in Section~\ref{sec:s3.1} are observed, we find that astrophysical uncertainties can lead to variations in the resulting neutrino flux of up to an order of magnitude. 
This level of uncertainty, however, remains much smaller than the large gap between the expected neutrino flux from 3C~279 and the sensitivity of current neutrino telescopes.
Its inclusion alongside the $W^{\pm}$ channel thus provides a more complete picture of leading electroweak processes in blazar jets.

\begin{figure}[t]
    \centering
    \includegraphics[width=0.7\linewidth]{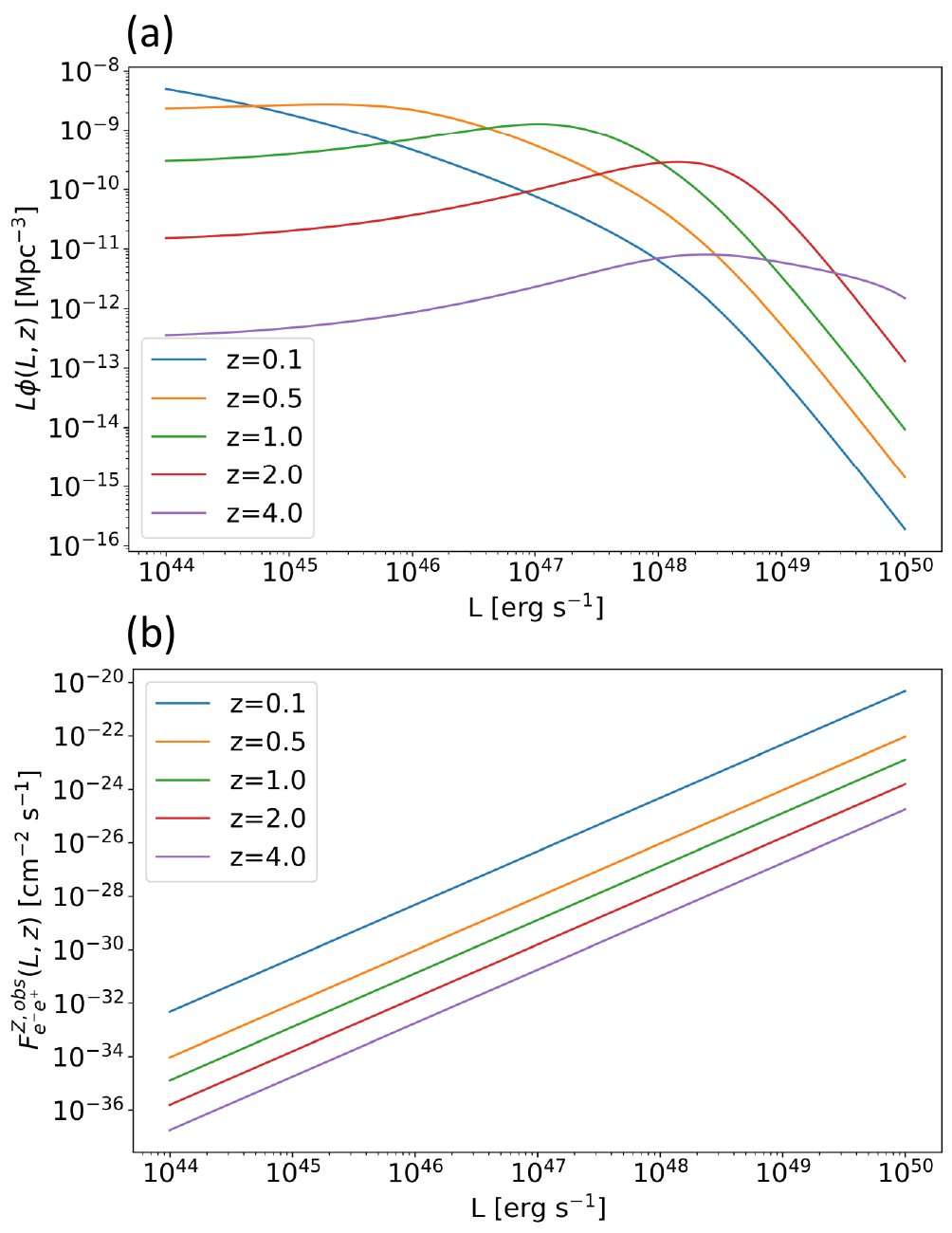}
    \caption{(a) Luminosity function of FSRQs and (b) corresponding observable diffuse neutrino flux from electron--positron interactions producing Z bosons, shown at different redshifts $z$. At $z > 2$, luminous sources ($L > 10^{48} {\rm erg~s^{-1}}$) dominate the contribution, while at $z < 1$ low-luminosity sources ($L < 10^{46} {\rm erg~s^{-1}}$) are more numerous but contribute less significantly to the flux.
    }
    \label{fig:f6}
\end{figure}

\begin{figure}[t]
    \centering
    \includegraphics[width=0.7\linewidth]{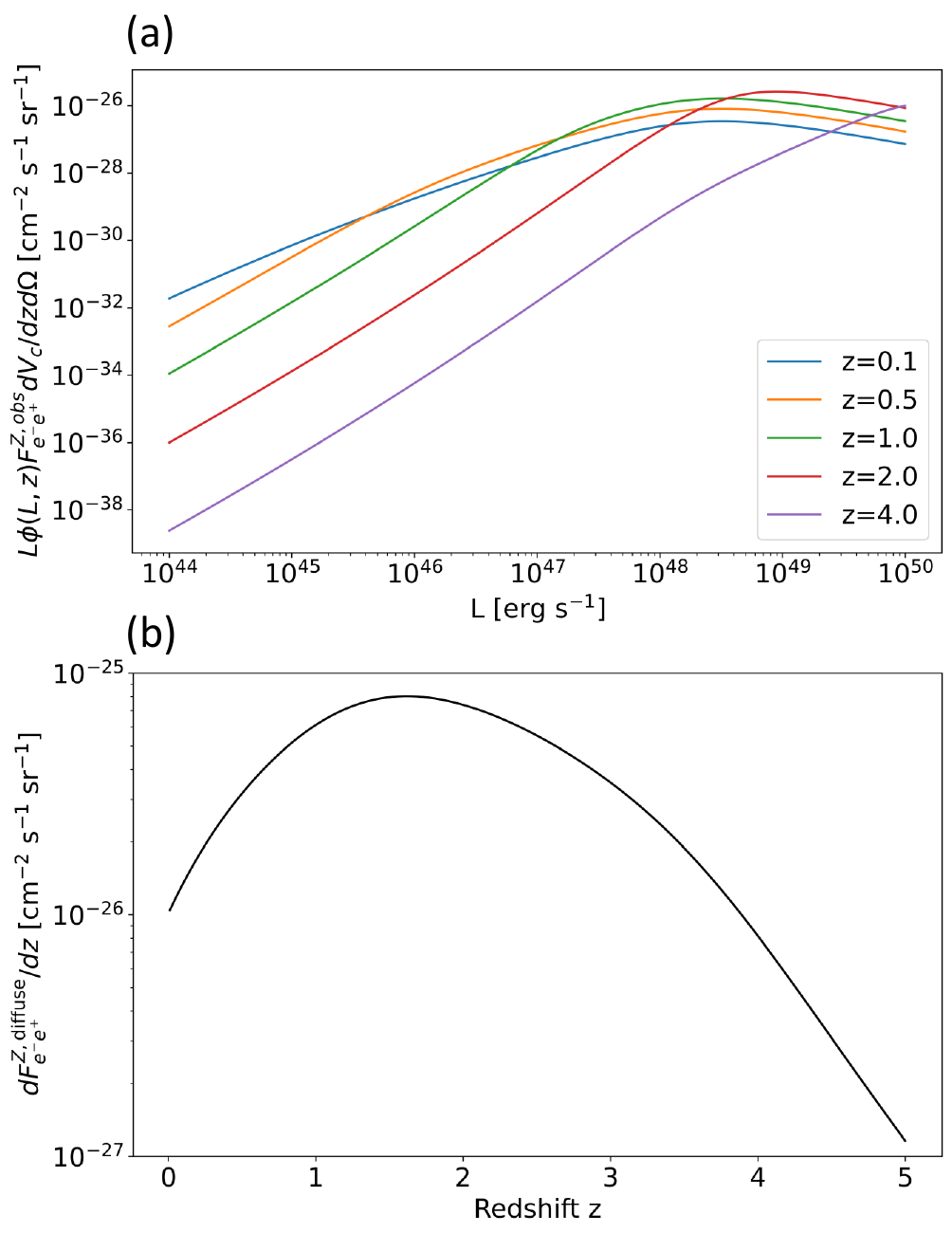}
    \caption{(a) Integrands of $F_{e^-e^+}^{Z,{\rm diffuse}}$ as functions of luminosity L for different redshifts $z$. At $z > 2$, the contribution is dominated by luminous sources with $L > 10^{48} {\rm erg~s^{-1}}$. (b) Differential flux contribution, $dF_{e^-e^+}^{Z,{\rm diffuse}}/dz$ integrated over luminosity, shown for $0.01 \leq z \leq 5$. The distribution exhibits a clear peak at $z \sim 1.5$, reflecting the cosmic evolution of the FSRQ population.
    }
    \label{fig:f7}
\end{figure}

\subsection{Diffuse neutrino flux from electron--positron collisions in FSRQ jets}
\label{sec:s3.3}

According to Sections \ref{sec:s3.1} and \ref{sec:s3.2}, the neutrino point-source detectability of resonant electroweak bosons ($W^{\pm}, Z$) from Flat Spectrum Radio Quasars (FSRQs) appears unlikely. 
This conclusion is further strengthened by the fact that the electron energy distribution in 3C~279 was modeled during its flaring state. The corresponding normalization would likely be lower in a quiescent state.
Therefore, the interaction rates derived for 3C~279 may be regarded as optimistic upper-limit estimates for individual sources. 
Although the contribution from each source is expected to be small, it is nevertheless meaningful to extrapolate the 3C~279-based interaction rates to the cosmological population of FSRQs, under the assumption that the rate approximately scales with the total nonthermal luminosity of the jet. 
Such extrapolation provides a benchmark estimate of the diffuse neutrino flux, while the luminosity function, source evolution, and cosmological effects are incorporated to account for the distribution and history of the FSRQ population.

We adopt the FSRQ luminosity function $\Phi(L,z)$ parameterized in~\cite{ajello2012} as a function of source luminosity $L$ and redshift $z$:
\begin{equation}
\Phi(L,z) = \frac{A}{\ln(10) \, L} \left[ \left(\frac{L}{L_*}\right)^{\alpha_1} + \left(\frac{L}{L_*}\right)^{\alpha_2} \right]^{-1} f_{\rm evol}(L,z),
\end{equation}
where the normalization $A = 3.4 \times 10^{-9} \, \mathrm{Mpc^{-3}}$, the break luminosity $L_* = 10^{48.3} \, \mathrm{erg \, s^{-1}}$, and the slopes $\alpha_1 = 0.21$, $\alpha_2 = 1.58$.  
The luminosity-dependent evolution factor $f_{\rm evol}(L,z)$ is
\begin{equation}
f_{\rm evol}(L,z) = \left[ \left( \frac{1+z}{1+z_c(L)} \right)^{p_1} + \left( \frac{1+z}{1+z_c(L)} \right)^{p_2} \right]^{-1},
\end{equation}
with
\begin{equation}
z_c(L) = z_c^* \left( \frac{L}{10^{48} \, \mathrm{erg \, s^{-1}}} \right)^{\kappa},
\end{equation}
and parameters $z_c^* = 1.47$, $\kappa = 0.21$, $p_1 = 7.35$, $p_2 = -6.51$.  
The event rate $\Gamma_{e^- e^+}^j(L)$ of electron--positron collisions producing electroweak bosons ($j = W^{\pm}, Z$) is assumed to scale with the blazar luminosity $L$ according to
\begin{equation}
\Gamma_{e^- e^+}^j(L) = \Gamma_{e^- e^+,0}^j \left(\frac{L}{L_0}\right)^2 ,
\end{equation}
where $L_0$ and $\Gamma_{e^- e^+,0}^j$ denote the luminosity and corresponding event rate of the reference source 3C~279.
In the optically thin regime the interaction rate scales as the product of the electron and positron number densities $n_{e^-} n_{e^+}$.
If the densities of both energetic electrons and positrons increase with the jet power traced by $L$ a quadratic dependence $\Gamma \propto L^2$ naturally follows.
This assumes that both electron and positron densities scale with the jet power in a similar manner, as expected in pair rich FSRQ jets where the lepton content is constrained by SED modeling.
This scaling therefore represents a physically motivated extrapolation to the blazar population reflecting the expectation that the density of high energy leptons and hence the collision rate grows with the total nonthermal power output of the jet which is commonly traced by the bolometric luminosity.
For a source at redshift $z$, the observed flux $F_{e^- e^+}^{j, \rm obs}(L,z)$ is given by
\begin{equation}
F_{e^- e^+}^{j, \rm obs}(L,z) \sim \frac{\Gamma_{e^- e^+}^j(L)}{4 \pi d_L^2(z) (1+z)},
\end{equation}
where $d_L(z)$ is the luminosity distance, and the factor $(1+z)$ accounts for cosmological time dilation.  
Figure~\ref{fig:f6} shows the luminosity functions and observable fluxes of electron--positron interactions producing $Z$ bosons at different redshifts $z$. 
For $z>2$, high-luminosity sources (i.e., $L > 10^{48} \, \mathrm{erg \, s^{-1}}$) dominate, and the observed fluxes from these energetic sources can exceed those from low-luminosity sources (i.e., $L < 10^{46} \, \mathrm{erg \, s^{-1}}$) at lower redshifts $z<1$. 
Despite the larger luminosity distance at $z>2$, the flux associated with electron--positron interactions is not negligible. 
This indicates that the diffuse flux cannot be attributed solely to nearby low-luminosity sources, but rather results from a balance between the luminosity function evolution and the increasing luminosity distance with redshift.

The total diffuse flux $F_{e^- e^+}^{j, \rm diffuse}$ observed on Earth from the cosmological population of FSRQs is computed by integrating over luminosity and redshift, accounting for the comoving volume element:
\begin{equation}
F_{e^- e^+}^{j, \rm diffuse} = \int_{z_{\rm min}}^{z_{\rm max}} dz \int_{L_{\rm min}}^{L_{\rm max}} dL \, \Phi(L,z) \, F_{e^- e^+}^{j, \rm obs}(L,z) \, \frac{dV_c}{dz d\Omega},
\end{equation}
where $dV_c/dz d\Omega$ is the comoving volume element per unit redshift and solid angle, obtained from the cosmological model \cite{aghanim2020}.
The redshift and source luminosity ranges are assumed to be $z=[0.01,5]$ and $L=[10^{44},10^{50}] \, \mathrm{erg \, s^{-1}}$, respectively, covering the typical luminosity and redshift distribution of FSRQs.  
To investigate the flux contribution from different cosmic epochs, the integrands of $F_{e^- e^+}^{Z, \rm diffuse}$ are shown in figure~\ref{fig:f7}-(a) as functions of luminosity $L$ at various redshifts $z$. As expected from the form of the luminosity function, sources at $z>2$ provide substantial contributions when their luminosities exceed $10^{48} \, \mathrm{erg \, s^{-1}}$.  
Integrating over luminosity yields the differential flux contribution, ${dF_{e^- e^+}^{Z, \rm diffuse}}/{dz}$, which is estimated as a function of redshift $z$ (figure~\ref{fig:f7}-(b)), exhibiting a clear peak around $z \sim 1.5$. 
This peak reflects the cosmic evolution of the FSRQ population, where the comoving number density of luminous sources reaches its maximum. 
At lower redshifts the volume density decreases, while at higher redshifts the increasing luminosity distance and declining source density reduce the overall contribution. 
Therefore, the differential flux peak around $z \sim 1.5$ represents a physically meaningful imprint of the evolutionary history of FSRQs.

Using the reference event rate for 3C 279, we obtain the following diffuse neutrino fluxes:
\begin{align}
F_{e^- e^+}^{W^\pm, \rm diffuse} &\sim 3.7 \times 10^{-36} \ \mathrm{cm^{-2} \, s^{-1} \, sr^{-1}},\\
F_{e^- e^+}^{Z, \rm diffuse} &\sim 1.5 \times 10^{-25} \ \mathrm{cm^{-2} \, s^{-1} \, sr^{-1}}.
\end{align}
The estimated fluxes represent the cumulative contribution of electron--positron annihilation into $W^\pm$ and $Z$ bosons across the cosmological population of FSRQs. 
Although the obtained values remain well below the diffuse astrophysical neutrino flux measured by current high-energy neutrino observatories ($\sim 10^{-12} \ \mathrm{cm^{-2} \, s^{-1} \, sr^{-1}}$ at 0.1--1~TeV) such as IceCube \cite[e.g.,][]{aartsen2017} or KM3NeT \cite[e.g.,][]{aiello2024}, they provide a theoretical benchmark for the role of leading (resonant) Standard Model processes in extreme astrophysical environments.  
This result highlights the connection between particle physics and astrophysics, demonstrating that even extremely rare high-energy interactions leave a subtle imprint on the cosmic radiation background. 
While the predicted flux is far below current detection capabilities, it provides a theoretical reference for diffuse electroweak signatures. 
In particular, when considering interactions confined within the jet blob, the resonant $Z$ boson production channel yields a flux that is many orders of magnitude smaller than the total diffuse astrophysical neutrino flux. 
Despite its extreme faintness, this fraction remains of theoretical interest as a probe of high-energy electroweak processes in FSRQ jets.
This interpretation continues to hold under the astrophysical uncertainties associated with modeling the electron energy distribution within the blob, including acceleration and energy losses.
Since the reaction rate may vary by about an order of magnitude as shown in figure~\ref{fig:f5}, such uncertainties are less significant when compared to the large gap between the diffuse neutrino flux expected from resonant $Z$ boson production and the sensitivity of currently available neutrino detectors.
In addition, the predicted fluxes can, in principle, be modified (for example, through enhanced cross sections) by incorporating new mechanisms of particle production in lepton collisions beyond the Standard Model.
The presented framework readily allows one to take into account such modifications.

\section{Summary and discussion}
\label{sec:s4}

In this study, we investigate the potential role of electron--positron collisions in the jet environments of Flat Spectrum Radio Quasars (FSRQs), specifically focusing on the leading production channels of electroweak bosons ($W^\pm$, $Z$). 
These interactions may contribute to the high-energy diffuse astrophysical neutrino background through processes such as the resonant annihilation into $Z$ bosons or the recently predicted Glashow resonance in $e^+ e^- \to W^\pm \rho(770)^{\mp}$.

We employ a one-zone leptonic model to describe the jet dynamics and electron distribution in the FSRQ 3C~279 during its flaring state, using the Fokker--Planck equation to model electron acceleration via shock processes and turbulence. 
This model allows us to estimate the reaction rates for the production of $W^\pm$ and $Z$ bosons from electron--positron collisions confined within the jet blob. 
The results show that while these processes are theoretically possible, the flux from such interactions is many orders of magnitude below the sensitivity limits of existing neutrino observatories such as IceCube, KM3NeT, or Baikal-GVD.
Consequently, the contribution from $Z$ boson production to the diffuse astrophysical neutrino flux is extremely small, far below previous estimates.

Despite the low detectability of these signals, the study highlights the importance of investigating electroweak interactions in extreme astrophysical environments. 
Since the modeling is based on the flaring state of 3C~279, the estimated interaction rates should be regarded as optimistic upper limits; quiescent states would likely yield even smaller contributions.  
Although the predicted fluxes remain beyond the reach of current observatories, future facilities such as IceCube-Gen2 or KM3NeT may begin to explore such scenarios, especially in synergy with multi-messenger observations of blazar flares. 
Our simplified one-zone framework provides a first theoretical benchmark, while more elaborate models including multi-zone dynamics, hadronic processes, or time-dependent variability will further refine the estimates. 

Beyond astrophysics, this work illustrates the potential of blazar jets as natural laboratories to probe electroweak processes at energies far exceeding those available in terrestrial accelerators, thereby bridging high-energy particle physics and cosmic neutrino astronomy. 
The presented framework also readily allows one to incorporate additional production channels, including scenarios beyond the Standard Model.

\acknowledgments
The authors thank S. V. Troitsky and D. S. Gorbunov for useful discussions. We also thank the referees for their careful reading of the manuscript and for their constructive comments and suggestions, which have significantly improved the quality and clarity of this work.




\end{document}